# Teaching HCI Design in a Flipped Learning M.Sc. Course Using Eye-Tracking Peer Evaluation Data

Michalis Xenos, Maria Rigou
Computer Engineering and Informatics Department, University of Patras, Patras, Greece
xenos@ceid.upatras.gr
rigou@ceid.upatras.gr

**Abstract:** This paper presents experiences from a flipped classroom M.Sc. course on Human-Computer Interaction (HCI). The students that finished successfully this course participated in twelve short (about two to three hours each) workshops, based on a flipped classroom model. Each workshop focused on a specific HCI activity, while before the workshops, a two-hour lecture was used to introduce the students in the flipped learning concept. This was the only lecture in this course, while all the rest of the educational material was offered to the students online before each workshop. Such material was mainly short lectures from the professor, in the form of videos uploaded in the course's YouTube channel and documents delivered using the university learning management system (LMS). For each workshop the students had to be prepared to participate, which was tested using brief quizzes before the start of specific workshops. The activity presented in this paper was the design and evaluation of an interactive system. The students were asked to form six groups comprising of three to four students each. Then a system's description, vague enough to stimulate creativity, was randomly assigned to each group. This activity presented in this paper was the longest activity of the entire course and it was conducted in four consequent workshops. The paper presents the setting of this experiment, the peer assessment method and the use of eye-tracking data collected and analysed to aid the students towards improving their design. The students created a working model of the system with limited functionality and improved this model using eye-tracking data from the peer evaluation of this model. The use of these data offered them the insight to improve their models and to undergo design changes. The paper presents samples of the progress made between various versions of the models and concludes presenting the preliminary positive results of the students' qualitative evaluation of this experiment.

**Keywords:** Flipped Classroom; Blended Learning; Human-Computer-Interaction; Eye-Tracking; Higher Education.

## 1. Introduction

Nowadays the boundaries between in campus and distance education are not as distinct as they used to be, especially in higher education. Today, higher education in-campus students have a plethora of online tools in hand, that vary from tools used for communication and socialising to pure e-learning tools. Using such tools transforms their learning experience from a typical in campus-based education model to a blended learning model (Garrison and Kanuka, 2004). In fact, we argue that most **campus-based learning today has changed into blended learning** since the use of online tools is not a novelty, but a commodity in almost all higher education programs.

Within this frame, a flipped classroom model (Bergmann and Sams, 2012) was used to teach Human-Computer Interaction (HCI) in an M.Sc. program. To the best of our knowledge, this is the first time a flipped classroom is used in an M.Sc. engineering course in Greece. The novelty of this approach is that we have used workshops focusing on the design of interaction and on the usability evaluation of these interaction designs, using real evaluation data derived from eye-tracking recordings of peer evaluation sessions. The participating students were able to follow all the steps of designing a system, starting from mock-up screens, creating a prototype and adding interactivity to their system. Furthermore, they switched roles and evaluated the systems developed by their peers and, finally, they used the evaluation data to improve their own prototypes. All these activities were organised into four workshops. In this paper, we present the setting of this course, the workshops details, and samples of the progress made between various versions of the models. Finally, we conclude by presenting the preliminary results of the students' qualitative evaluation of this experiment.

The rest of the paper is structured as follows. Section 2 presents a brief literature review of similar educational models for teaching HCI and for using eye-tracking data for the evaluation of interaction design, as well as within a learning process. Section 3 presents the course and the "Software Quality and Human-Computer Interaction Laboratory" that hosted the peer evaluation, while section 4 presents the activities that took place in the four workshops in detail, illustrated with examples from students' designs. Finally, section 5 summarizes the conclusions from applying this educational process and presents the preliminary results of the students' qualitative evaluation of this process.

## 2. Literature review

While blended learning is not something new, the more the technology infiltrates into everyday practices the more all campus-based learning programs are moving towards blended learning. Therefore, although blended learning was initially considered as the mean to move learners from traditional classrooms to e-learning in small steps making change easier to accept (Driscoll, 2002), nowadays blended learning is a common practice for most campus-based programs, especially in higher education. This is happening because higher education institutes have the infrastructure to combine web-based technologies and the appropriate pedagogical approaches with face-to-face lectures. Today, most campus-based higher education courses offer the course material online, handle submission and assessment of students work through an LMS, facilitate an online community through e-fora and messages, even scheduling online office hours. Furthermore, blended learning is not mostly about the tools and the approaches, but it is about the concept that learning is not something related to a one-time event (i.e. during the lecture), but a continues process (Singh, 2003). This concept is acknowledged in most higher education institutes today, leading to the implementation of various blended models (Drossos et al., 2008).

One of the most effective blended learning strategies is flipped learning (Bergmann and Sams, 2012) where the lecture is moved outside of the classroom and inside the classroom the students perform activities. These activities are mostly group-based collaborative activities following learner-centred learning theories (Vygotsky, 1980). This flipped classroom, therefore, uses a learner-centred model in which the activities into the classroom explore topics in greater depth that students have already studied online. On the one hand, the problem of flipped learning is that it requires a lot more effort from the educator, compared to the traditional lecture preparation, since it is an expansion of the curriculum, rather than a mere re-arrangement of activities (Bishop and Verleger, 2013). On the other hand, if used appropriately and if the educator took the time and effort to prepare the online lectures videos and design the in-classroom activities, flipped learning is a powerful educational method.

Using the flipped learning strategy in an HCI course was an obvious choice since HCI is a multi-discipline field that requires collaborative learner-centred activities. A definition of HCI is "*Human-computer interaction is a discipline concerned with the design, evaluation, and implementation of interactive computing systems for human use and with the study of major phenomena surrounding them*" (Hewett et al., 1992) and, therefore, teaching HCI is a difficult practice that has to evolve in response to changes in the technological landscape (Churchill et al., 2013). Following our choice of flipping the classroom the creation of all the required online material (video lectures) and the design of the activities took an entire year of preparation. The educational material used for the course was mostly videos from the professors and a limited number of selected online short videos, enhanced with reading material and, in some cases, software tools. These online videos were developed using detailed principles (Pierrakeas et al., 2003) and following specific quality guidelines.

One of the activities we have selected for this course was the use of peer evaluation, based on eye-tracking data since eye-tracking allows recording and analysing detailed eye gaze data, offering insight on how users spontaneously react to visual stimuli and overall interaction design. The basic measure is the gaze point, which equals one raw sample captured by the eye-tracker. Fixations aggregate a series of gaze points and represent a period in which the eyes are locked towards a specific point. Between fixations, there are quick movements called saccades. The ordered set of fixations points (depicted by circles) connected by saccades (depicted by lines) is called a 'gazeplot' (or 'scanpath', or 'gazetrail'). A 'heatmap' is another visualization offered, where colours or opacity vary with the density of the number of fixations or their duration. Eye-tracking metrics can also be extracted based on a sub-region of the displayed stimuli (e.g., specific images, blocks of text, calls to action, etc.), defined as an area of interest (AOI). AOIs can be defined during the analysis process as the most relevant areas of the stimuli.

Despite the limitations of eye-tracking technology (sensitive to head positioning, thick glasses or contact lenses and inability to capture peripheral vision) there is significant research concerning this technology or based on it, in numerous application domains. By adequate interpretation of eye-tracking data, scientists can measure attention, interest, and arousal, and reach interesting conclusions for human behaviour research applied in a variety of fields such as Cognitive Studies and Education, Psychology, Medicine, Neurology, UX and HCI, Marketing, Engineering and more. In the cognitive studies and educational research, several visual psychologists have concluded that eye movement is an objective indicator for thoroughly monitoring and analysing the cognitive processes of learners (Baker and Loeb, 1973, Rayner, 1998, Salvucci and Anderson, 2001). Sung and Tang (2007) found that eye movement such as gaze time is a reliable index to observe cognitive processing in sentence reading, while Sanders and McCormick (1987) concluded that more than 80% of human beings manage to process cognitive information through visual processes. Thus, eye movement is an essential source of information in the cognitive processes and has been used in several studies to examine learning processes, study visual attention as well as social interaction in various learning settings. Research has indicated that eye-tracking can contribute to studying information gathering, problem-solving, learning strategies, interaction patterns between teachers and students or among students, as well as the effectiveness of various educational resources (Koc-Januchta et al., 2017, Kohlhase et al., 2017, Lai et al., 2013, Lin et al., 2016, Lin et al., 2017, Merkley and Ansari, 2010, Rosch and Vogel-Walcutt, 2013, Tien et al., 2014). In the domain of Computer Science teaching, Obaidellah et al. (2018) surveyed the use of eye-tracking in assessing the underlying cognitive processes of programming and Busjahn et al. (2015) conducted an eye-tracking study of the way students read programming code compared to natural language text and also surveyed the use of eye-tracking in computing education (Busjahn et al., 2014).

## 3. The HCI course and the equipment used for the activity

The activity presented in this paper took place on the “Computer Science and Engineering” M.Sc. program and in particular in the course “Human's Interaction with Computers, Robots and Smart Devices”. The evaluation took place in the “Software Quality and Human-Computer Interaction Laboratory”. The course and the laboratory equipment are presented in brief in this section.

### 3.1 The HCI course

The course “Human's Interaction with Computers, Robots and Smart Devices” is part of the “Computer Science and Engineering” M.Sc. program. This M.Sc. program is an 18-month (3 semesters) program offering 90 ECTS (30 ECTS per semester) available at the Computer Engineering and Informatics department of the University of Patras. The students participating in this program are required to complete 12 courses and a thesis. The “Human's Interaction with Computers, Robots and Smart Devices” course is a core (i.e. compulsory) course for the students of the division of “Computer Software” and an elective course for the students of the other two divisions (“Hardware and Computer Architecture” and “Applications and Foundations of Computer Science”) of this M.Sc. program. This course was offered to students for the first time on the academic year 2016-2017. During the second semester, starting on February 2017 to May 2017 of this academic year, 24 students registered in this course and 22 of them finished it successfully. Since the two students that haven’t finished haven’t participated in the activities presented in this study, the course population is 22 students and is called hereinafter in this paper “the students”. These students represent a typical Computer Science (CS) population, 4 female and 18 male, with mean age around 27 years.

The course outline of the “Human's Interaction with Computers, Robots and Smart Devices” included design and evaluation techniques for contemporary interactive devices such as cars, appliances, smartphones, robotic devices and multimodal computer interfaces. The students that successfully completed this course participated in 12 short (2-3 hours each) workshops, having to perform a specific activity in each one of them. Activities that are not presented in this paper included teamworking, debates, card-based activities, using specific software, and presentations. The first lecture of the course was a short (about two hours) lecture, which was used to introduce the students in the flipped learning concept. No further lectures were given, and the rest of the educational material was offered to students to study it before each workshop, following a typical flipped classroom model. This material was mostly short video lectures that were uploaded in the course’s YouTube channel, as well as documents and tools that were offered through the university’s LMS, which allowed collecting learning analytics (Koulocheri and Xenos, 2013) for the material used. The use of videos was not limited to the lectures only. Short videos were used to deliver the professors’ comments on the students’ projects between workshops. Students were required to be prepared to participate in the workshops and in some of them, a short quiz was introduced before the workshop started, to evaluate their level of preparation.

The activity ‘from mock-up screens to interaction design’, presented in detail in section 4, was the longest activity of the entire course and it was conducted in four consequent workshops, from the 4th workshop to the 7th workshop of this course. These workshops are called workshop A, B, C and D for the rest of this paper, in order not to confuse the reader with the ones that are not presented hereinafter. The other activities included debates on open HCI issues at the 1st and the 2nd workshop, using the IoT Toolkit (Mora et al., 2017) and cards to stimulate creativity at the 3rd workshop, improving the efficiency of the user interface based on the keystroke level model (Card et al., 1980) and using the KLM-FA tool (Karousos et al., 2013) at the 8th and the 9th workshops, using the Greek version (Katsanos et al., 2012) of the standard usability scale (SUS) questionnaire (Brooke, 1996) and focus groups to evaluate their prototypes' usability at the 10th workshop, and students' presentations of selected Human-Robot Interaction (HRI) papers at the 11th and 12th workshops.
To the best of our knowledge, this is one of the first courses offered using the flipped classroom model in an M.Sc. course in Greece and the first on HCI that combines such a variety of activities in an engineering M.Sc. course.

### 3.2 The laboratory and the equipment used

The "Software Quality and Human-Computer Interaction Laboratory" lab at the Computer Engineering and Informatics Department is configured in two separate adjacent spaces with visual contact, a testing room and an observation room. The laboratory offers equipment for collecting eye-tracking data and physiological signals to measure users' stress (Liapis et al., 2015, Liapis et al., 2017). The test for this assignment was conducted in the testing room using the Tobii T120 Eye-tracker integrated in a 17-inch TFT monitor. Test scenarios were set up in Tobii Studio software which was also used for data analysis and visualization. Users were recorded on video and were encouraged to express freely their thought and opinion concerning their current task (think-aloud).

## 4. From mock-up screens to interaction design

The activity presented in this paper is the iterative design and evaluation of an interactive system. This activity was the longest one of the course and was conducted in 4 consequent workshops (A, B, C and D) presented in this section. During the preparation for these four workshops, and while studying the educational material that was available online for them, the students were asked to form groups of 3 to 4 persons. They were totally independent on how to form the groups and they could use the course's e-forum and messaging tools, as well as any other means of communication they choose. The only thing requested from them was that the names of the students of each group to be available before the start of the first workshop of this series. They were also informed that studying the corresponding material was essential for the successful completion of the workshop, so they should make sure that no members of their team would show up unprepared, since this would have a negative impact on the entire team's performance.
Following our instructions, the students formed 4 groups of four persons and 2 groups of three persons, while we prepared six folders each one having a system's description which was vague enough to stimulate creativity. To balance the groups' effort, two systems that according to the professors needed less effort were prepared for the 3-persons groups, while four systems estimated to require more effort were prepared for the 4-persons groups.
The systems were selected to address a variety of users with diverse needs, to allow the design of interaction based on various modalities and to emphasise on several aspects of the interaction, such as efficiency, error prevention, perceived satisfaction, etc. Therefore, the six folders included the following system descriptions:

1. A game for small children that would be played inside the classroom in pairs, where children could learn basic arithmetic operations.
2. A system supporting an anaesthesiologist during a surgery that would require as input the drugs and their dosage, using multimodal interactions and it could monitor the patient and report to the anaesthesiologist during the operation.
3. A smartphone application for small children that could control a teleoperated toy car, but that would be probably played by their parents as well.
4. A system for the captain of a large ferryboat that allows the operator to open and close hatches on various car docks, while using security mechanisms to prevent human errors.
5. A system for elderly people that would serve as information desk at a hospital they visit for a routine check, after reading their social security card, to inform them about the options they have and to schedule appointments.
6. A subsystem of the previous system that provides directions on how to find their next destination (e.g. the office they should go next).

The systems 1 and 6 were the ones aimed for the smaller groups. The folders included details on what is required from each system but did not reveal any information about how this is going to be designed, or what features could be included, or what modalities might be the important ones. The students were told that they could consider all contemporary technologies to their disposal, but the goal of the activity is to create something that the users would find attractive, useful and efficient, rather than impress them with the use of state-of-the-art technology.

## 4.1 Workshop A

This was a three-hour workshop that started with students selecting their systems. Since all systems descriptions were hidden into folders of the matching size, the selection among groups of the same size was random. Firstly, the two small (3-persons) groups chose their folders and then the rest of the groups did the same. Then, the students divided into groups and they to discuss about their system and to define the basic personas. When all teams were ready, the class was regrouped, and each team presented their personas to the class, where they received feedback from the professor and their peers. Following this, the students separated once again in groups and they had to design basic mock-up screens and user interactivity using pen and paper. Figure 1 shows two mock-up designs for two of these systems created during the workshop. For this activity, they also received feedback when the class was once again regrouped.

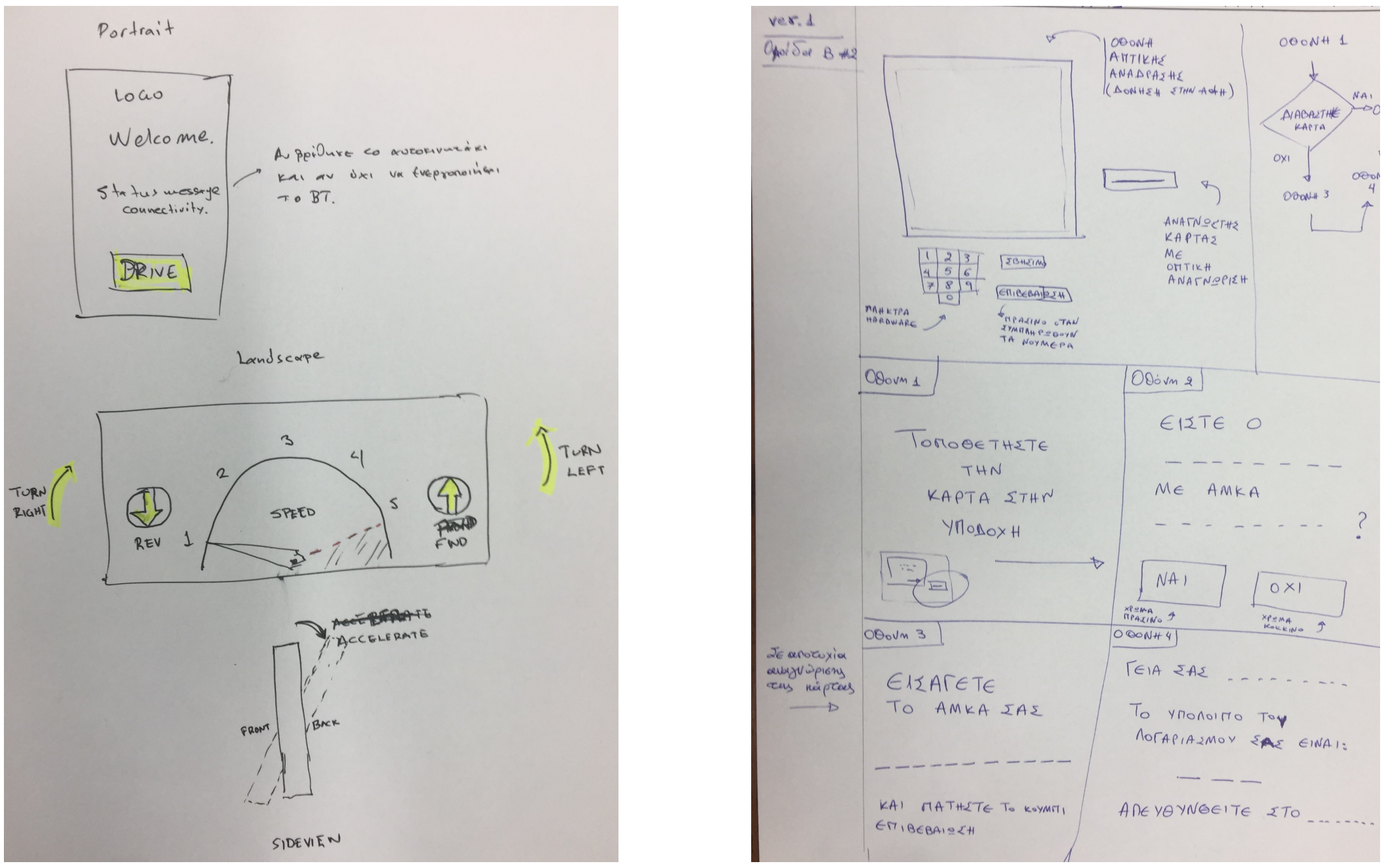

**Figure 1:** Two samples of the mock-up designs for system no. 3 and 5 respectively

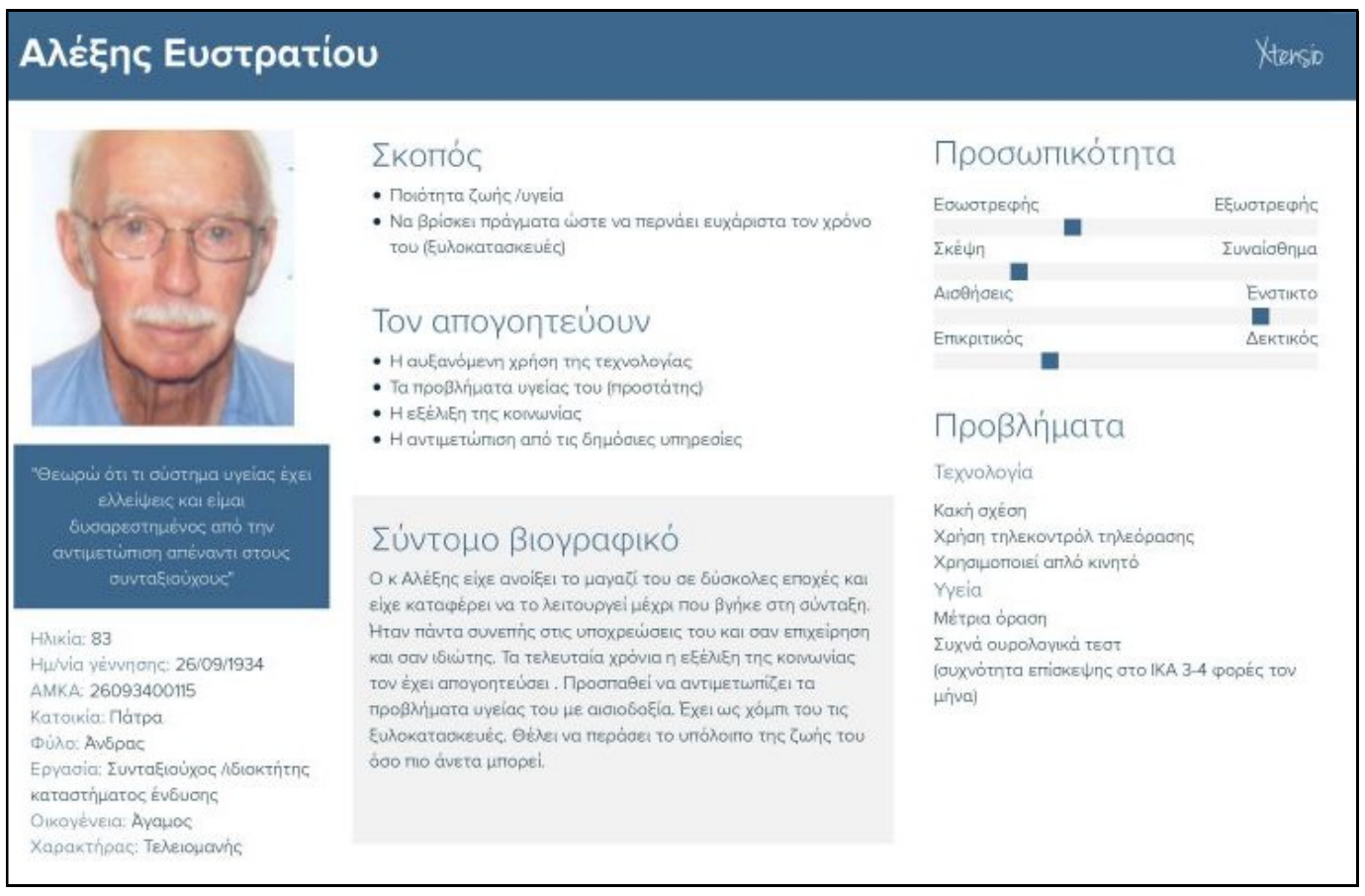

**Figure 2:** A persona created by the group working on system no. 6

After this workshop and while preparing for the next one, the students had to present their personas online and to design a rapid prototype (with adequate interactivity) based on their mock-up screens and the feedback they had received. A sample of a persona submitted by a group is shown in Figure 2, where the students have enriched their personas with personal details, a short CV, personality details and problems related to the technology and the system they investigated. Finally, the students had to provide detailed scenarios of the interaction and to submit all these online before the beginning of the next workshop.

### 4.2 Workshop B

Workshop B was a 3-hour workshop where each student used all the other teams' prototypes, based on the prepared scenarios, using eye-tracking. Following workshop B, students received the data from the eye-tracking process for their prototype and the outlined scenario (individual user and grouped heatmaps and gazeplots, as well as statistical data on AOIs students specified) and were also given the option to return to the eye-tracker to watch recordings and access all available metrics for more thorough interpretations and better redesign decisions. Figure 3 depicts one test scenario in Tobii Studio, Figure 4 presents the heatmap and Figure 5 the gazeplot (with mouse clicks) corresponding to user gaze data when asked to turn off the sound on the mobile car control application (users should click on the Settings icon placed at the top centre part of the screen).

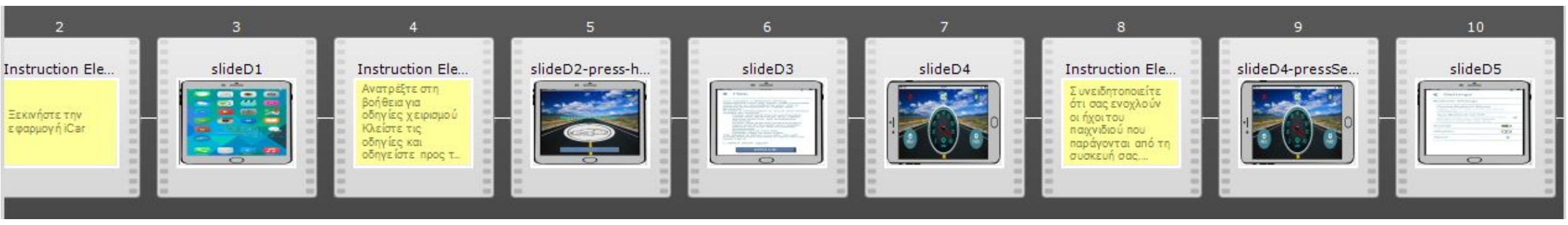

**Figure 3:** Test scenario in Tobii studio for system no. 3

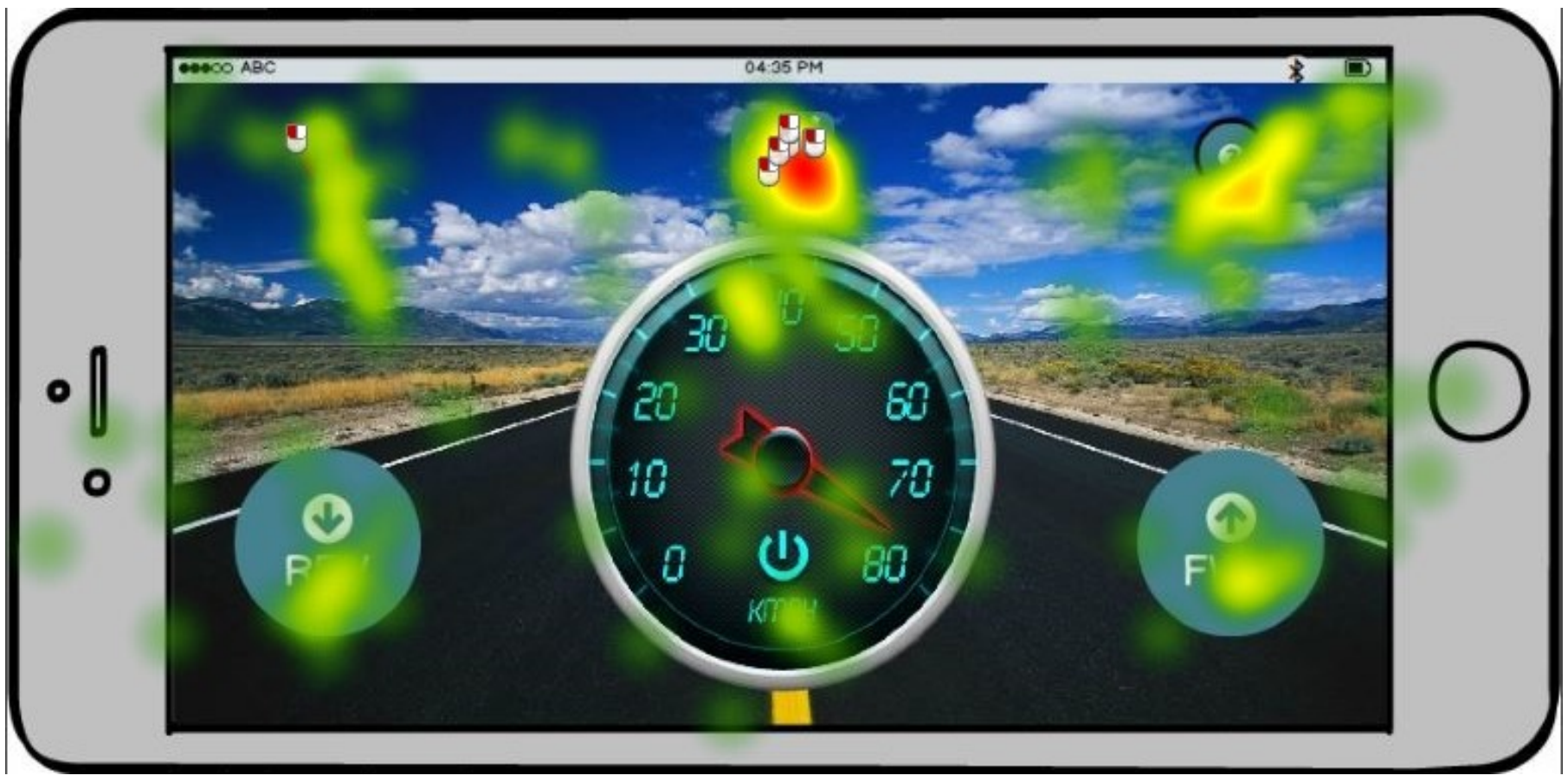

**Figure 4:** Heatmap annotated with mouse clicks (all users) for system no. 3

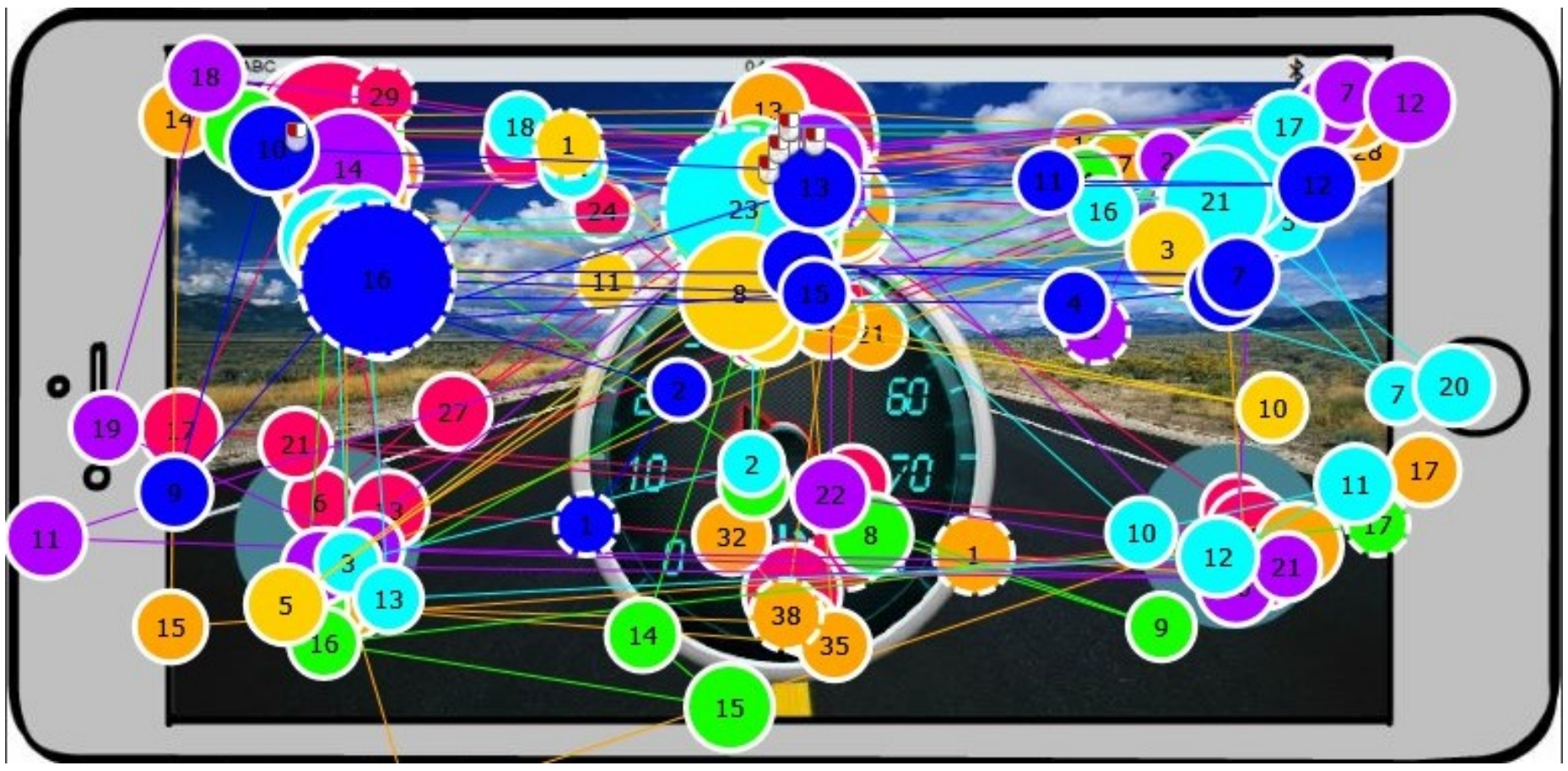

**Figure 5:** Gazeplot annotated with mouse clicks (all users) for system no. 3

## 4.3 Workshop C

This was a two-hour workshop, where the members of each group had the chance to discuss with the evaluators their comments and their actions during the interaction with the system based on the scenarios they used. Therefore, students played both the roles of the designer and the evaluator, since all of them participated in the design of their system and in the evaluation of other systems. This workshop involved a lot of students moving from their group to other groups, allowing them to have a global view of how all teams approached the design of their assigned system. After this workshop, and while preparing for the next one, the students used all these data to redesign their prototype and to improve their design.

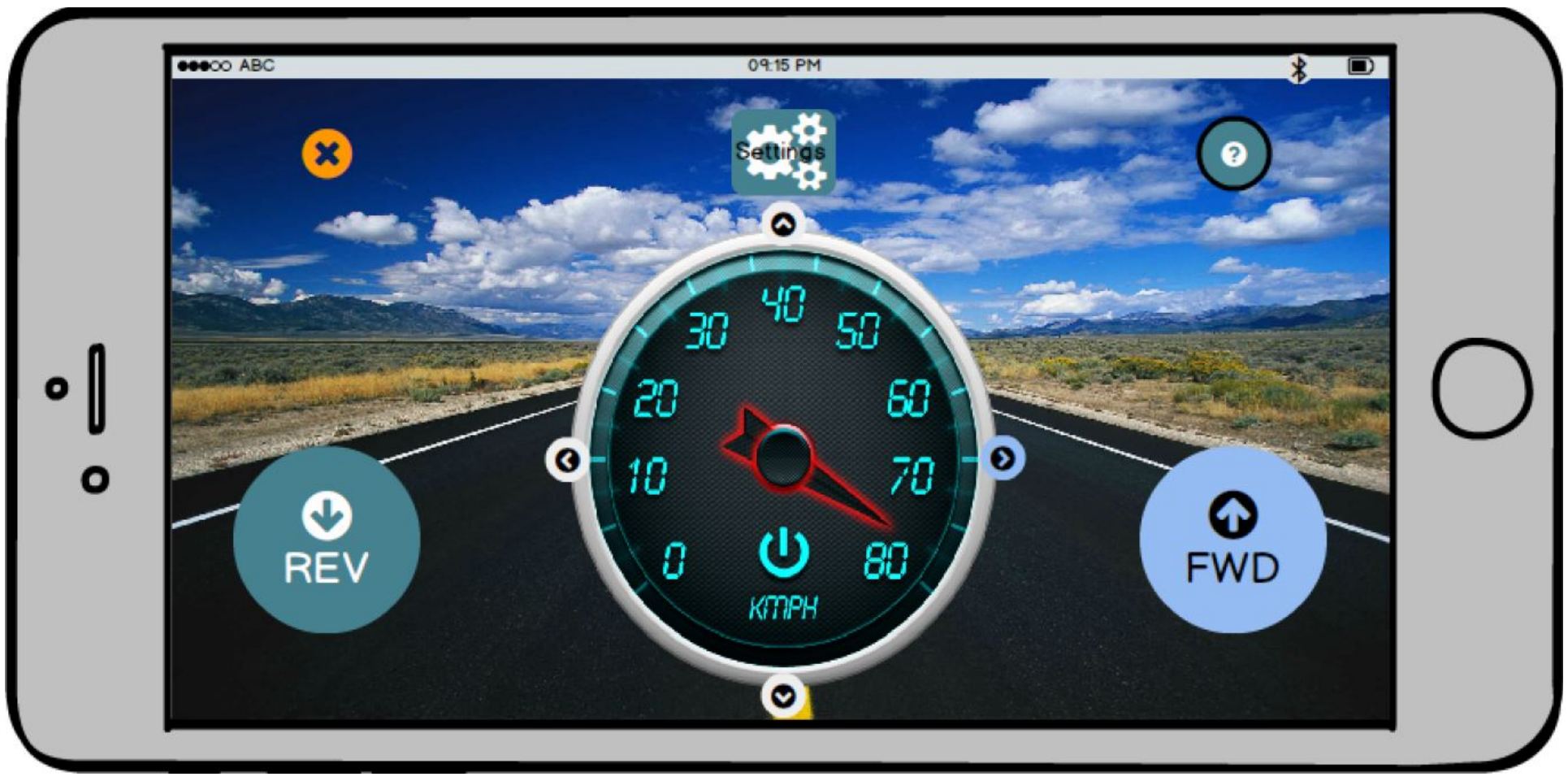

**Figure 6:** A sample of a final design, following the results of the evaluation

## 4.4 Workshop D

This was a three-hour workshop where all student groups had about 25 minutes each (including questions) to present their systems design and to discuss with the professors and peers the concept, the process and the improvements they made throughout the entire activity. Figure 6 presents a final system as presented by the students in this workshop (this design was based on the mock-up depicted in Figure 1, left side). In this example, as observed by both visualizations in Figure 4 and Figure 5 most users managed to locate the required option (Settings) but their visual attention was also drawn by option Help, as well as Exit (colored in a visually dominating red) and one user clicked on Exit rather than Settings. This led students to the assumption that the recognisability of Settings option needed to be enhanced so a label was added in the redesign of the game central screen (Figure 6). Moreover, students decided to use a less vivid colour for Exit.

# 5. Results and conclusions

The students that participated in this course had the opportunity to work both as designers of an interactive system and as usability evaluators. This setting serves well the learning objectives of the course, as students gained valuable practical and analytical experience and they should be able to design eye-tracking testing sessions from now on according to their needs. In addition, their experience with evaluating the designs of other students provides them with good and bad examples of UI design which also contributes to their HCI learning.

Furthermore, students responded enthusiastically to how this course was conducted. On the formal online assessment tool, used by the University of Patras, where students evaluate all courses they attend anonymously, this course had scored from 4.00 to 4.59 in each evaluation category (in a typical Likert 1 to 5 scale) with the lowest score being related to course difficulty (4.00) and the highest score related to the content delivery and collaboration with the students. This is a very high score compared to similar scores of other courses. Furthermore, the qualitative comments were also very positive, i.e.: "*…it was the first time that I have participated in such a well-organised course…*", "*…I loved the activities since most of these were both useful and fun to participate…*". Finally, this work is not without limitations. To effectively measure the educational value of these activities we should be able to compare learning gain measured in comparison to a traditional lecture-based classroom. A between-subjects experiment to investigate this is a future goal.

Another future goal is to offer to the participating students more measurements (e.g. physiological measurements).